\documentclass{article}

\usepackage{arxiv}

\usepackage[utf8]{inputenc} 
\usepackage[T1]{fontenc}    
\usepackage{hyperref}       
\usepackage{url}            
\usepackage{booktabs}       
\usepackage{amsfonts}       
\usepackage{nicefrac}       
\usepackage{microtype}      
\usepackage{lipsum}		
\usepackage{graphicx}
\usepackage{natbib}
\usepackage{doi}
\usepackage{xcolor}
\usepackage{color}
\usepackage{multirow}
\usepackage{subfigure}
\usepackage{caption}
\usepackage[resetlabels]{multibib}

\hypersetup{
colorlinks=true,
linkcolor=blue
}

\title{BioMedGPT: Open Multimodal Generative Pre-trained Transformer for BioMedicine}

\author{%
    \textbf{Yizhen Luo}$^{1}$\thanks{\quad Supported by PharMolix Inc.}\enspace, 
    \textbf{Jiahuan Zhang}$^{1}$\footnotemark[1]\enspace, 
    \textbf{Siqi Fan}$^{1}$\footnotemark[1]\enspace, 
    \textbf{Kai Yang}$^{1}$\footnotemark[1]\enspace, 
    \textbf{Yushuai Wu}$^{1}$\footnotemark[1]\enspace, 
    \textbf{Mu Qiao}$^{2}$\thanks{\quad Corresponding author}\enspace, 
    \textbf{Zaiqing Nie}$^{1, 2}$\footnotemark[2]\\
    Institute for AI Industry Research (AIR), Tsinghua University$^{1}$\\
    PharMolix Inc.$^{2}$ \\
    \texttt{yz-luo22@mails.tsinghua.edu.cn,}\enspace \texttt{mqiao@pharmolix.com}\\
    \texttt{\{zhangjiahuan, fansiqi, yangkai, zaiqing\}@air.tsinghua.edu.cn}\\
}
\date{}

\renewcommand{\headeright}{}
\renewcommand{\undertitle}{}
\renewcommand{\shorttitle}{}

\hypersetup{
pdfkeywords={First keyword, Second keyword, More},
}

\begin{document}

\maketitle

\begin{abstract}
Foundation models (FMs) have exhibited remarkable performance across a wide range of downstream tasks in many domains. Nevertheless, general-purpose FMs often face challenges when confronted with domain-specific problems, due to their limited access to the proprietary training data in a particular domain. In biomedicine, there are various biological modalities, such as molecules, proteins, and cells, which are encoded by the language of life and exhibit significant modality gaps with human natural language. In this paper, we introduce BioMedGPT, an open multimodal generative pre-trained transformer (GPT) for biomedicine, to bridge the gap between the language of life and human natural language. BioMedGPT allows users to easily ``communicate'' with diverse biological modalities through free text, which is the first of its kind. BioMedGPT aligns different biological modalities with natural language via a large generative language model, namely, BioMedGPT-LM. We publish BioMedGPT-10B, which unifies the feature spaces of molecules, proteins, and natural language via encoding and alignment. Through fine-tuning, BioMedGPT-10B outperforms or is on par with human and significantly larger general-purpose foundation models on the biomedical QA task. It also demonstrates promising performance in the molecule QA and protein QA tasks, which could greatly accelerate the discovery of new drugs and therapeutic targets. In addition, BioMedGPT-LM-7B is the first large generative language model based on Llama2 in the biomedical domain, therefore is commercial friendly. Both BioMedGPT-10B and BioMedGPT-LM-7B are open-sourced to the research community. In addition, we publish the datasets that are meticulously curated for the alignment of multi-modalities, i.e., PubChemQA and UniProtQA. All the models, codes, and datasets are available at \url{https://github.com/PharMolix/OpenBioMed}.


\end{abstract}


\section{Introduction}

Foundation models (FMs) are large AI models trained on enormous amounts of unlabelled data through self-supervised learning. FMs such as ChatGPT \citep{chatgpt}, Bard \citep{bard}, and Chinchilla \citep{Hoffmann2022TrainingCL} have showcased impressive general intelligence across a broad spectrum of tasks. Most recently, Meta introduced Llama2~\citep{Touvron2023Llama2O}, a family of pre-trained and instruction-tuned large language models (LLMs), demonstrating prevailing results over existing open-source LLMs and on-par performance with some of the closed-source models based on human evaluations in terms of helpfulness and safety. 

However, these general-purpose FMs are typically trained on internet-scale generic public datasets, and their depth of knowledge within particular domains is restricted by the lack of access to proprietary training data. To solve this challenge, domain-specific FMs are becoming more prevalent, and attracting tremendous attention to their widespread applications in many fields. For instance, BloombergGPT~\citep{bloomberggPT} is a large-scale GPT model for finance. ChatLaw~\citep{cui2023chatlaw} fine-tunes Llama on a legal database and shows promising results on downstream legal applications.

Recently, research works about FMs in the biomedical domain have emerged. BioMedLM, an open-source GPT model with 2.7 billion parameters, is trained exclusively using PubMed abstracts and the PubMed Central section from the Pile data~\citep{gao2020pile}. PMC-Llama \citep{pmc}, a biomedical FM fine-tuned from Llama-7B with millions of biomedical publications, has demonstrated superior understanding of biomedical knowledge over general-purpose FMs. Biomedical FMs excel in grasping human language but struggle with comprehending diverse biomedical modalities, including molecular structures, protein sequences, pathways, and cell transcriptomics. Biomedicine, akin to human and computers, has its own distinct ``languages'', like the molecular language, which employs molecular grammars to generate molecules or polymers~\citep{pmlr-v202-guo23h}. Recent research efforts have been devoted to harnessing large-scale pre-trained language models for learning these multi-modalities. To bridge the modality gap, BioTranslator~\citep{Xu2023} develops a multilingual translation framework to translate text descriptions to non-text biological data instances. In particular, BioTranslator fine-tunes PubMedBERT \citep{pubmedbert} on existing biomedical ontologies and utilizes the resulting domain-specific model to encode textual descriptions. BioTranslator bridges different forms of biological data by projecting encoded text data and non-text biological data into a shared embedding space via contrastive learning. 

The inherent laws of nature and the evolution of life, dominated by the organization and interaction of atoms and molecules, employ \textit{the language of life} to constitute the first principle in biomedicine. Specifically, the language of life describes how information is encoded in logical ways to ensure proper functionalities underlying all of life. For example, the nucleic acids of DNA encode genetic information while the amino acid sequence contains codes to translate that information into protein structures and functions. In addition, proteins are the key building blocks in any organism, from single cells to much more complex organisms such as animals. On the other hand, humans have accumulated extensive biomedical knowledge over centuries, which are usually described in natural language and represented in the forms of knowledge graphs, text documents, and experimental results. There are still many territories yet to be explored by humans in order to understand the fundamental codes, i.e., \textit{the language of life}. For example, as of June 2023, UniProtKB/TrEMBL \citep{UniProt} records more than 248M proteins, but only 0.2\% of them are well studied and manually annotated by human experts. Recently, specialized AI models like ESM-2 \citep{ESM2} have shown promise in mining the substantial uncharted regions in biomedicine. However, these black-box models are incapable of providing scientific insights with human interpretable language. More recently, GPT-4 \citep{openai2023gpt4} demonstrates great success in comprehending not only natural language but also more structured data, such as graphs, tables, and diagrams, illuminating an opportunity to bridge the gap between \textit{the language of life} and natural language, and therefore revolutionize scientific research.

\begin{figure}[htp]
  \centering
  \makebox[\textwidth][c]{\includegraphics[width=1.1\linewidth]{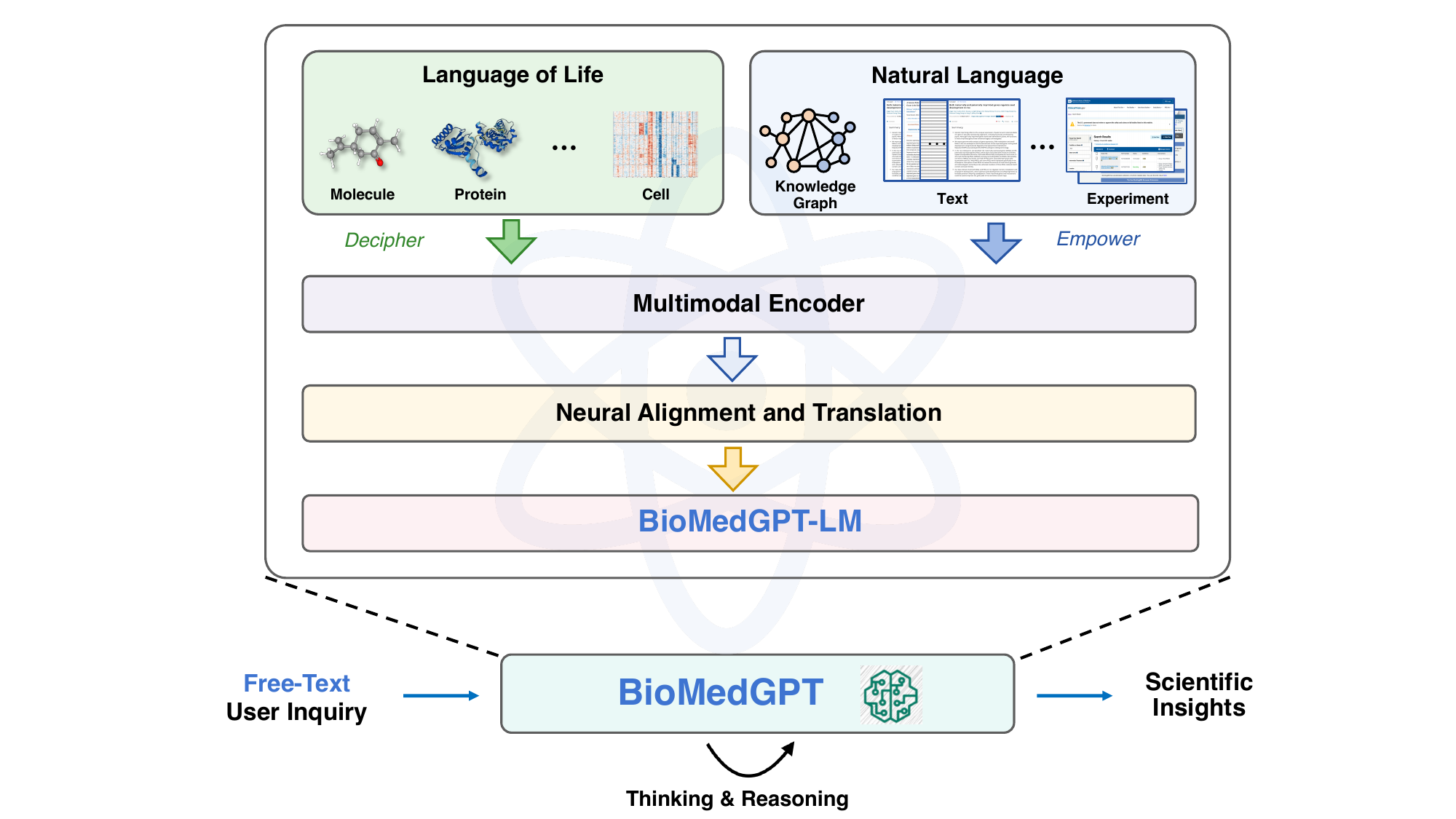}}
  \captionsetup{font={small}}
  \caption{The architecture of BioMedGPT.}
  \label{biomedgpt-u}
\end{figure}

To achieve this overarching and ambitious goal, we develop BioMedGPT, a novel framework to bridge the language of life and human natural language using large-scale pre-trained language models. Figure \ref{biomedgpt-u} shows the overview of BioMedGPT. We build a biomedical language model, BioMedGPT-LM, by fine-tuning existing general-purpose LLMs with large-scale biomedical corpus. Built on top of BioMedGPT-LM, BioMedGPT is designed to unify the language of life, encoding biological building blocks such as molecules, proteins, and cells, as well as natural languages, describing human knowledge in the forms of knowledge graphs, texts, and experimental results. Alignment and translation between different biological modalities and human natural language are performed via neural networks and self-supervised learning techniques. Compared with prior work, BioMedGPT enjoys the benefits of larger GPT type of LLMs and a substantial amount of untapped texts in the biomedical research field. Within the unified feature space of different modalities, BioMedGPT can flexibly support free-text inquiries, digest multimodal inputs, and versatilely power a wide range of downstream tasks without end-to-end fine-tuning.

We open-source BioMedGPT-10B, a model instance of BioMedGPT, unifying texts, molecular structures, and protein sequences. BioMedGPT-10B is built upon BioMedGPT-LM-7B, which is fine-tuned from the recently released Llama2-Chat-7B model with millions of biomedical publications. We leverage independent encoders to encode molecular and protein features and project them into the same feature space of the textual modality via multi-modal fine-tuning. BioMedGPT-10B enables users to upload biological data of molecular structures and protein sequences and pose natural language queries about these data instances. This capability can potentially accelerate the discovery of novel molecular structures and protein functionalities, thus catalyzing advancements in drug development.

Our contributions are summarized as follows:

\begin{itemize}
    \item We introduce BioMedGPT, a novel framework to bridge the language of life and human natural language via large-scale generative language models. 
    
    \item We demonstrate the promising performance of BioMedGPT-10B, which is a model instance of BioMedGPT, on the biomedical QA, molecule QA, and protein QA tasks. Through fine-tuning, BioMedGPT-10B outperforms or is on par with human and significantly larger general-purpose foundation models on biomedical QA benchmarks. In addition, 
    its capability on the molecule QA and protein QA tasks shows great potential to power many downstream biomedical applications, such as accelerating the discovery of new drugs and therapeutic targets. BioMedGPT-10B is open-sourced to the community. We also publish the datasets that are curated for the alignment of multi-modalities, i.e., PubChemQA and UniProtQA. 

    \item BioMedGPT-LM-7B, the large language model used in BioMedGPT-10B, is the first generative language model that is fine-tuned from Llama2 with an extensive biomedical corpus. BioMedGPT-LM-7B is commercial friendly and is open-sourced to the community. 
    
\end{itemize}

The remaining of the paper is organized as follows. Section 2 provides an overview of BioMedGPT. In Section 3, we present BioMedGPT-10B, which is a 10B foundation model in the BioMedGPT family, including BioMedGPT-LM-7B and molecule QA and protein QA modules. Experimental results and analysis are reported in Section 4. We describe the limitations of our work in Section 5. Finally, we conclude and discuss future work in Section 6.

\section{An Overview of BioMedGPT}
In Figure~\ref{biomedgpt-u}, we present an overview of BioMedGPT, which serves as a biomedical brain. The cognitive core, BioMedGPT-LM, is a large language model developed through incremental training on an extensive biomedical corpus, inheriting the benefits of both the emergent abilities of LLMs and domain-specific knowledge. BioMedGPT-LM serves not only as a linguistic engine that enables free-text interactions with humans but also acts as a bridge connecting various biomedical modalities. BioMedGPT is endowed with the ability to comprehend and reason over diverse biological modalities encompassing molecules, proteins, transcriptomic, and more, through the feature space alignment. In addition, we also utilize heterogeneous expert knowledge from knowledge graphs, text documents, and experimental results to further enhance the biomedical knowledge of BioMedGPT.


We develop a feature fusion approach, which encodes multimodal data with independent pre-trained encoders and aligns their feature spaces with that of the natural language through neural alignment and translation methods. We have demonstrated the effectiveness of this approach in our prior work, MolFM \citep{molfm}, which is a molecular foundation model that enables joint representation learning on molecular structures, biomedical texts, and knowledge graphs. MolFM achieves state-of-the-art performance on a spectrum of multimodal tasks, such as molecule-text retrieval, molecule captioning, and text-to-molecule generation. Thus, we believe that BioMedGPT can harvest from both the readily available uni-modal foundation models and the powerful generalization capability of language models. BioMedGPT enhances the cross-modal comprehension and connections between natural language and diverse biological modalities. Users can flexibly present inquiries in various formats, encompassing but not limited to, text, chemical structure files, SMILES, protein sequences, protein 3D structure data, and single-cell sequencing data. With extensive training on these biomedical data, BioMedGPT can provide users with valuable scientific insights. 

In the subsequent section, we introduce a model instance of BioMedGPT which is primarily focused on the joint comprehension of molecular structures, protein sequences, and biomedical texts.

\begin{figure}[htp]
  \centering
  \makebox[\textwidth][c]{\includegraphics[width=1.05\linewidth]{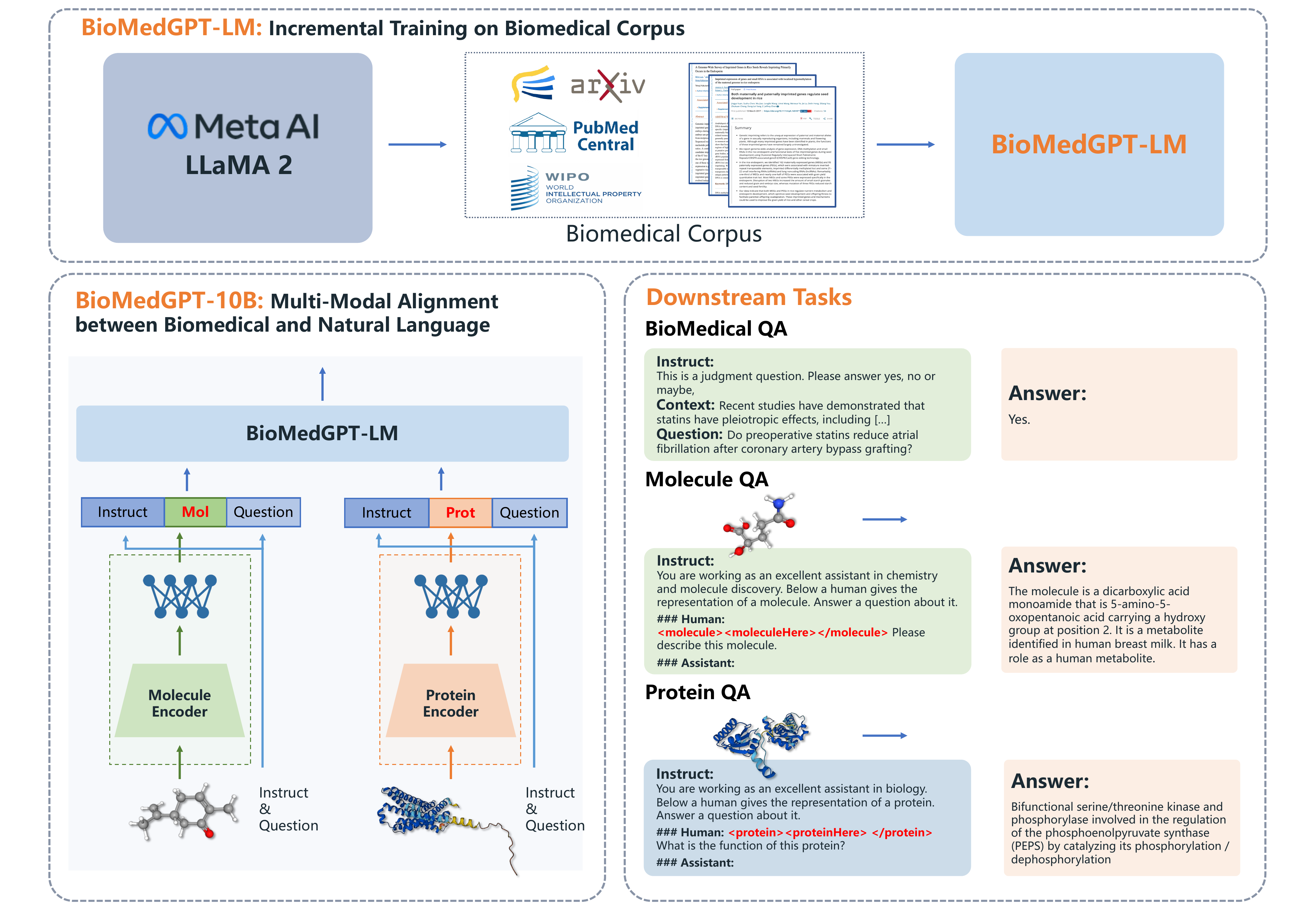}}
  \captionsetup{font={small}}
  \caption{The overview of BioMedGPT-10B. \textbf{BioMedGPT-LM} is the large language model of BioMedGPT, which serves as a cognitive core to jointly comprehend various biological modalities through natural language. In BioMedGPT-10B, the parameter size of the large language model is 7B. \textbf{BioMedGPT-10B} adopts GraphMVP \citep{GraphMVP} as the 2D molecular graph encoder, ESM2-3B \citep{ESM2} as the protein sequence encoder, and conducts feature space alignment via a neural network adaptor. BioMedGPT can be applied to many \textbf{multimodal downstream tasks} such as biomedical QA, molecule QA, and protein QA.} 
  \label{biomedgpt}
\end{figure}

\section{BioMedGPT-10B: Aligning Molecules, Proteins, and Natural Language}

As shown in Figure \ref{biomedgpt}, we develop BioMedGPT-LM-7B, a large language model specialized in biomedicine, through incremental training with biomedical literature on top of Llama2-7B-Chat. We then build BioMedGPT-10B by aligning 2D molecular graphs, protein sequences, and natural language in a unified feature space. We choose molecules and proteins because they are basic biomedical elements. It is worth noting that the aforementioned molecule and protein encoders can be easily replaced with other suitable and well-performing encoders.
We detail the architecture and training process of BioMedGPT-10B in the following subsections.

\subsection{BioMedGPT-LM-7B: Incremental training on large-scale biomedical literature}

Recently, Llama2 and Llama2-chat, open-sourced by Meta, have attracted great research attention owing to their outstanding capabilities in the general domain. To exploit the advantages of Llama2's emergent abilities as well as biomedical knowledge from scientific research, we perform incremental training on Llama2-Chat-7B with extensive biomedical documents from S2ORC~\citep{lo-wang-2020-s2orc}. 

\paragraph{Data Processing}
We select biomedical-related literature from S2ORC to fine-tune Llama2. The S2ORC dataset comprises an extensive collection of $81.1$ million English academic papers spanning various disciplines. We meticulously extract 5.5 million biomedical papers from this dataset using PubMed Central (PMC)-ID and PubMed ID as criteria. After removing articles without full text and those with duplicate IDs, we attain a refined dataset of 4.2 million articles. Our subsequent processing involves the removal of author information, reference citations, and chart data from the main body of each article. The remaining text is then partitioned into sentence-based chunks. Each of these chunks is further tokenized by the Llama2 tokenizer, culminating in a substantial assemblage of over 26 billion tokens highly pertinent to the field of biomedicine.

\begin{figure}[htpb]
  \centering
  \vspace{-0.4cm}
  \makebox[\textwidth][c]{\includegraphics[width=0.6\linewidth]{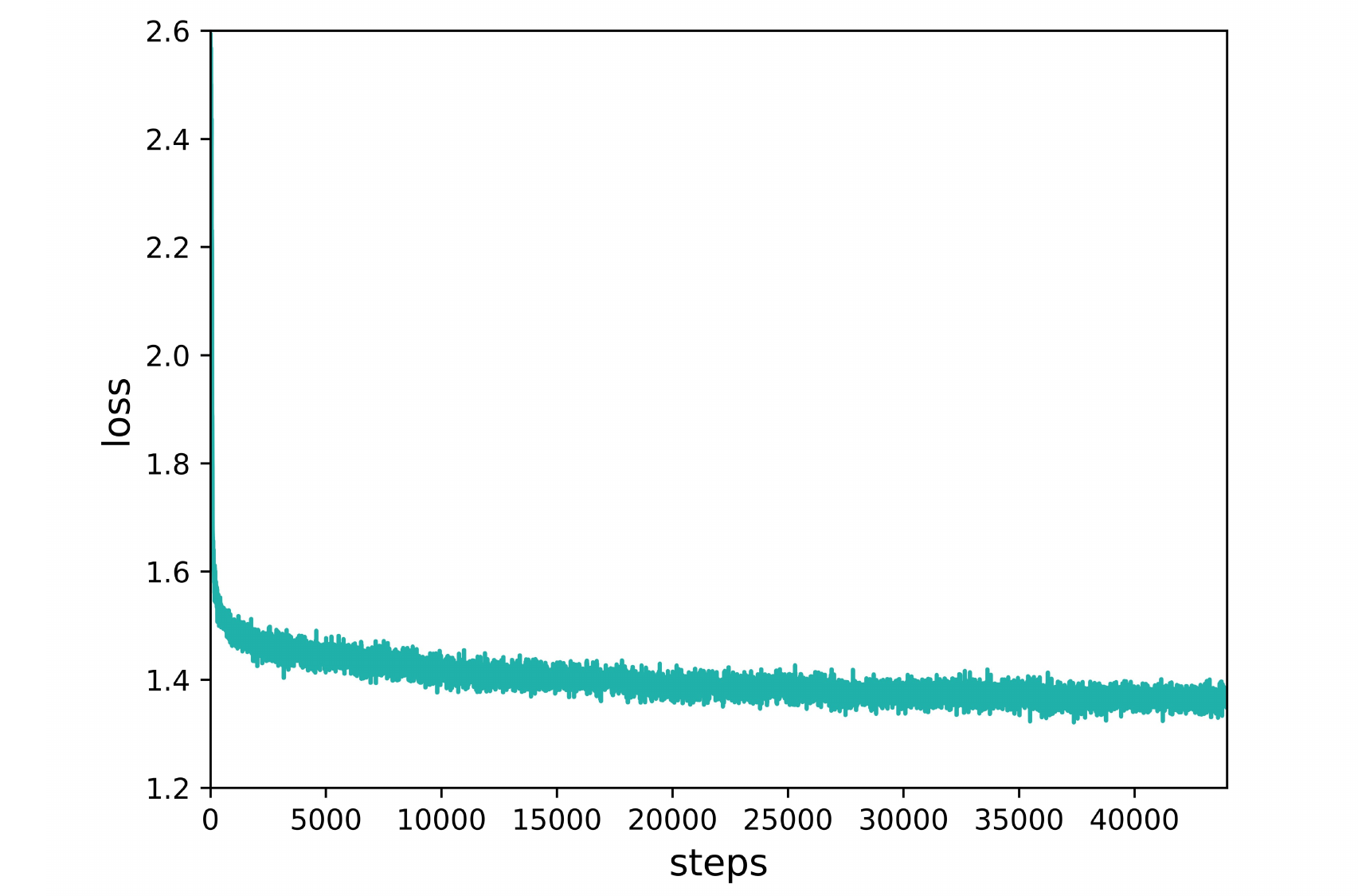}}
  \captionsetup{font={small}}
  \caption{The training loss for BioMedGPT-LM-7B.}
  \label{loss_figure}
\end{figure}

\paragraph{Fine-tuning Details}
 The fine-tuning utilizes a learning rate of $2 \times 10^{-5}$, a batch size of 192 \footnote{The batch size in this paper refers to the total batch size across all the GPUs.}, and a context length of 2048 tokens. We adopt the Fully Sharded Data Parallel (FSDP) acceleration strategy alongside the bf16 (Brain Floating Point) data format. To tackle the memory challenge, we leverage gradient checkpointing \citep{chen2016training} and flash attention \citep{dao2022flashattention}. 
 We utilize an autoregressive objective function. The training loss is shown in Figure \ref{loss_figure}. Notably, the loss exhibits a consistent and progressive decrease after the 44,000 steps, indicating the model converges effectively.

\subsection{Multimodal alignment between molecules, proteins, and natural language} 

\label{mma}
BioMedGPT-10B is composed of a molecule encoder, a protein encoder, a large language model (i.e., BioMedGPT-LM-7B), and two modality adaptors. 
In order to exploit the strong capabilities of existing unimodal models, we leverage off-the-shelf checkpoints to initialize the molecule and protein encoder. The molecule encoder is a 5-layer GIN \citep{Xu2018HowPA} with 1.8M parameters pre-trained by GraphMVP \citep{GraphMVP}, which shows promising results in comprehending 2D molecular graphs. The protein encoder is ESM-2-3B \citep{ESM2}, a 36-layer transformer \citep{vaswani2023attention} specialized in processing protein sequences. The output features of each atom of the molecule encoder and the output features of each residue of the protein encoder are projected to the feature space of BioMedGPT-LM-7B with independent modality adaptors composed of a fully-connected layer.

To build the connections between molecular and protein structures with natural language, we perform multimodal fine-tuning, which involves answering questions with regard to a given molecule or protein. As shown in Table \ref{prompt}, we design prompt templates to help BioMedGPT-LM understand the context more accurately in a role-play manner. The \texttt{<moleculeHere>} and \texttt{<proteinHere>} symbolics represent the aligned molecular and protein features, where each atom of a molecule and each residue of a protein is considered as a token. \texttt{\{text\_input\}} is populated by questions in the training set, and \texttt{\{text\_output\}} is populated by answers to calculate the auto-regressive loss. 

\paragraph{Data Proccessing} 
\label{Alignment data}
The multimodal alignment is performed on two large-scale datasets that we curate, namely, \textbf{PubChemQA} and \textbf{UniProtQA}. We publicly release these datasets to facilitate future research.
\begin{itemize}
    \item PubChemQA consists of molecules and their corresponding textual descriptions from PubChem \citep{PubChem}. It contains a single type of question, i.e., \texttt{please describe the molecule}. We remove molecules that cannot be processed by RDKit \citep{rdkit} to generate 2D molecular graphs. We also remove texts with less than 4 words, and crops descriptions with more than 256 words. Finally, we obtain 325, 754 unique molecules and 365, 129 molecule-text pairs. On average, each text description contains 17 words.
    \item UniProtQA consists of proteins and textual queries about their functions and properties. The dataset is constructed from UniProt \citep{UniProt}, and consists 4 types of questions with regard to functions, official names, protein families, and sub-cellular locations. We collect a total of 569, 516 proteins and 1, 891, 506 question-answering samples. The data was randomly divided into training, validation, and test sets at a ratio of $8:1:1$. The multi-modal fine-tuning is performed on the training set of ProteinQA.
\end{itemize}

We design the following prompt to organize the molecular or protein data with the text data as feature-ordered input for the LLM.

\begin{table}[h]
    \caption{Prompt for organizing multi-modality data entry.}
    \centering
    \label{prompt}
    \resizebox{0.85\textwidth}{!}{
        \begin{tabular}{c|p{0.8\textwidth}}
            \toprule
            \textbf{Modality} & \textbf{Prompt} \\
            \midrule
            Molecule & \texttt{You are working as an excellent assistant in chemistry and molecule discovery. Below a human gives the representation of a molecule. Answer a question about it.} \\
            & \#\#\# \texttt{Human: <molecule><moleculeHere></molecule> \{text\_input\}.} \\
            & \#\#\# \texttt{Assistant: \{text\_output\}} \\ 
            \midrule
            Protein & \texttt{You are working as an excellent assistant in biology. Below a human gives the representation of a protein. Answer a question about it.} \\
            & \#\#\# \texttt{Human: <protein><proteinHere></protein> \{text\_input\}.} \\
            & \#\#\# \texttt{Assistant: \{text\_output\}}\\ 
            \bottomrule
        \end{tabular}
    }
\end{table}

Following mPLUG-owl \citep{ye2023mplugowl}, we freeze the parameters of BioMedGPT-LM and optimize the parameters of the molecule encoder, protein encoder, and modality adaptors to save the computational cost and avoid catastrophic forgetting. We conduct fine-tuning using these two datasets.


\section{Experiment}
In this section, we substantiate BioMedGPT-10B's capability to jointly understand and model the language of life and natural language with a series of experiments. We present three question-answering tasks, namely biomedical QA, molecule QA and protein QA, to comprehensively evaluate the biomedical knowledge that our model encompasses. In the following sections, we will introduce the datasets and experimental results for each task. Additionally, we showcase the generated results in protein QA.

\subsection{Biomedical QA} 

Biomedical QA involves answering free-text questions in the biomedical domain, which challenges the professional level of language models. The task serves as a means to evaluate if BioMedGPT-10B can understand biomedical terminologies and reason over complex contexts like a human expert.

\paragraph{Dataset}

We evaluate BioMedGPT-10B on three public multiple-choice question answering benchmarks within the biomedical domain, i.e., \textbf{USMLE}~\citep{usmle}, \textbf{MedMCQA}~\citep{medmcqa}, and \textbf{PubMedQA}~\citep{pubmedqa}.

The details of the datasets used in biomedical QA are summarized below.

\begin{itemize}
    
    \item \textbf{USMLE} is a real-world medical QA dataset collected from the United States Medical License Exams with 11, 451 professional multiple-choice questions. We follow the official data splits with a training and test ratio $8:1$. 
    \item \textbf{MedMCQA} is a popular medical QA dataset covering diverse healthcare topics and subjects. More than 194k high-quality medical multiple-choice questions from real-world entrance exams are collected. The training set contains 182,822 QA pairs, and the test set consists of 4,183 pairs.
    \item \textbf{PubMedQA} is a QA dataset in biomedical research. It aims at answering research questions with yes/no/maybe according to the corresponding contexts collected from the PubMed abstracts. Following ~\citep{pmc}, we use the manually curated subset, PQA-A, which contains 211,269 QA pairs as our training set. The evaluation is done on the labeled subset, PQA-L, which consists of 1,000 QA pairs.
    
\end{itemize}

\paragraph{Experiment Setup} We fine-tune the language model of BioMedGPT-10B on the training sets of PubMedQA and MedMCQA which consists of a similar amount of data. Then we perform in-domain (ID) evaluation on the test sets of two datasets, and out-of-domain (OOD) evaluation on USMLE without additional fine-tuning on its training set. The model is fine-tuned for 3 epochs with a learning rate of $2\times 10^{-5}$ and a batch size of 512. 

\paragraph{Evaluation} 
We compare the QA performance of BioMedGPT-10B with zero-shot InstructGPT \citep{ouyang2022training}, ChatGPT \citep{chatgpt}, Llama \citep{touvron2023llama}, Llama2-Chat \citep{Touvron2023Llama2O} as well as fine-tuned Llama, Llama2-Chat and PMC-Llama \citep{pmc}. We also report the accuracy of passing the test and the accuracy of human experts. The in-domain results are displayed in Table \ref{tab:qa-id}. We observe that Llama2 outperforms the previous Llama model, and Llama2-Chat further boosts the accuracy since QA tasks can be considered as a single-round dialogue. Additionally, fine-tuning language models brings substantial improvements. Benefiting from the generalization power of Llama2-Chat and incremental fine-training with large-scale biomedical corpus, BioMedGPT-10B achieves state-of-the-art results on both MedMCQA and PubMedQA. Notably, on PubMedQA, the prediction accuracy of BioMedGPT-10B is on par with human experts, demonstrating its great potential to serve as a professional biomedical assistant.

We also report the OOD performance of BioMedGPT-10B and baseline models on USMLE in Table~\ref{tab:qa-id}. Compared with other LLMs of the same parameter size, BioMedGPT-10B shows outstanding performance, surpassing the best model (i.e., Llama2-Chat) by a notable 5.1\% margin. Besides the closed-source ChatGPT, BioMedGPT-10B is the only model that achieves more than 50\% accuracy on USMLE.

\begin{table}[h]
    \caption{Performance (accuracy, \%) comparison for ID (In-Domain) and OOD (Out-Of-Domain) evaluation. All the reported results with * are referred from LMFlow \citep{lmflow}.}
    \centering
    \begin{tabular}{c|c|ccc}
    \toprule
    Method &   Setting &  MedMCQA(ID) &  PubMedQA(ID) &  USMLE(OOD) \\
    \midrule

    Human (pass)*   &  \multirow{2}*{Manual} & -      & 60.0 & 50.0  \\
    Human (expert)* &                         & 90     & 78.0 & 87.0 \\  
    \midrule
    InstructGPT* &  \multirow{5}*{Zero-shot} & 44.0  & 73.2 & 46.0  \\
    ChatGPT*       &                         & 44.7   & 63.9 & 57.0 \\
    Llama*      &                         & 24.3   & 5.2 & 27.1 \\
    Llama2    &                         & 30.6   & 3.7  & 27.2 \\
    Llama2-Chat &                         & 35.5   & 21.9 & 34.4 \\ 
    \midrule
    Llama \citep{pmc} &  \multirow{3}*{Fine-tuning} & 48.2 & 73.4 & 44.6  \\
    Llama2-Chat &                        & 48.3   & 75.5  & 45.3    \\ 
    PMC-Llama \citep{pmc} &                & 50.5   & 69.5 & 44.7    \\ 
    \midrule    
    BioMedGPT-10B &  Fine-tuning         & \bf51.4   & \bf76.1   & \bf50.4    \\
    \bottomrule
    
    \end{tabular}
    \label{tab:qa-id}
\end{table}

\subsection{Molecule QA} 

Molecule QA involves generating a text response given a specific molecule and a text query over its properties. This task serves as a means to evaluate the ability of our model to translate between natural language and the language of molecules.

\paragraph{Dataset} 
We incorporate ChEBI-20 \citep{edwards-etal-2021-text2mol}, a dataset incorporating 33, 010 molecules from ChEBI (Chemical Entities of Biological Interest) with high-quality textual descriptions extracted from PubChem as our benchmark. We adopt \texttt{"Please describe this molecule."} as the question, and challenges language models to generate a textual description based on the input structure. We follow the initial split with a training, validation, and test ratio $8:1:1$.

\paragraph{Experiment Setup} We evaluate the model performance including BLEU, ROUGE, and METEOR following \citep{MolT5}. We fine-tune BioMedGPT-10B using the same prompts as Table \ref{prompt}. The fine-tuning procedure consists of 50 epochs with a learning rate of $7\times 10^{-5}$ and a batch size of 24, and we freeze the parameters of the language model to reduce computation cost. We also evaluate the performance of Llama2-7B-Chat by filling the molecule SMILES string into the prompt \texttt{<moleculeHere>} portion.



\paragraph{Evaluation} 
Table~\ref{tab:M-C} shows the experimental results on the test set of the ChEBI-20 dataset. Comparing with ChatGPT and Llama2-7B-Chat, the substantial performance enhancement of BioMedGPT-10B demonstrates the importance of multi-modal alignment and translation. Despite of their potency, general-purpose LLMs lack the capacity to directly interpret molecule language. 


The cross-modal alignment mechanism of BioMedGPT-10B not only facilitates the model's comprehension of the semantic aspects of molecular language but also leverages its robust language modeling capability to effectively translate molecular language into human language, providing a flexible tool for experts and non-experts to grasp the fundamental knowledge of a molecule. 

\begin{table}[htbp]
    \caption{Performance comparison on molecule QA.}
    \small
    \centering
    \begin{tabular}{c|c|ccccccc}
        \toprule
           Method & Alignment & BLEU-2 & BLEU-4 & ROUGE-1 & ROUGE-2 & ROUGE-L & MEATOR\\
        \midrule
        ChatGPT \citep{li2023empowering} & \multirow{2}*{ w/o} & 0.103 & 0.050 & 0.261 & 0.088 & 0.204 & 0.161\\
        Llama2-7B-Chat   &                          & 0.075 & 0.009 & 0.184 & 0.043 & 0.142 & 0.149\\ 
        \midrule
        BioMedGPT-10B       &  w/ & \textbf{0.234} & \textbf{0.141} & \textbf{0.386} & \textbf{0.206} & \textbf{0.332} & \textbf{0.308} \\
        \bottomrule
    \end{tabular}
    \label{tab:M-C}
\end{table}

\subsection{Protein QA}
Protein QA involves generating a text response to a query about a given protein, which is formulated as an amino acid sequence. Protein QA requires the capability to jointly decypher the properties and  functions from protein sequences and grasp semantics within questions.

\paragraph{Dataset} We perform the evaluation on the test set of UniProtQA. Details of our dataset are presented in Section \ref{Alignment data}.

\paragraph{Experiment Setup} 
We evaluate the performance with the same evaluation metrics as Molecule QA. Since the multimodal alignment is done by fine-tuning the training set of UniProtQA, we perform testing directly with the same prompt in Table \ref{prompt}. For comparison, we implement two baselines: (1) Vanllina Llama2-7B-Chat, where we fill the prompt \texttt{<proteinHere>} portion with raw amino acid sequence and perform zero-shot testing. (2) Llama2-7B-Chat with alignment, where we substitute the language model with Llama2-7B-Chat and perform the same multi-modal alignment as introduced in Section \ref{mma}.
\begin{table}[h]
    \caption{The Performance of BioMedGPT-10B on Protein QA.}
    \small
    \centering
    \begin{tabular}{c|c|ccccccc}
        \toprule
            & Alignment & BLEU-2 & BLEU-4 & ROUGE-1 & ROUGE-2 & ROUGE-L & MEATOR  \\
        \midrule
        Llama2-7B-Chat & w/o  & 0.019 & 0.002 & 0.103 & 0.060 &  0.009 & 0.052 &     \\
        Llama2-7B-Chat & w/  & 0.344 & 0.313 & 0.705 & 0.711 & 0.593 & 0.707 &     \\
        \midrule
        BioMedGPT-10B    & w/ & \bf0.571 & \bf0.535 & \bf0.743 & \bf0.759 & \bf0.622 & \bf0.754 &  \\
        \bottomrule
    \end{tabular}
    \label{P-C}
\end{table}

\paragraph{Evaluation} 
The results of Protein QA are shown in Table \ref{P-C}. Comparisons between two baselines demonstrate that protein sequences exhibit a prominent modality gap with natural language. The challenges in comprehending proteins lead to out-of-order and meaningless responses of general-purpose language models. While aligning protein language with human language serves as an effective solution to address this issue, we observe a significant performance gain of BioMedGPT-10B over Llama2-7B-Chat with alignment, especially in terms of BLEU metrics. This indicates that our incremental training on biomedical corpus can help language models better comprehend and generate professional terminologies.

\begin{figure}[htpb]
  \centering
  \vspace{-0.3cm}
  \makebox[\textwidth][c]{\includegraphics[width=1\linewidth]{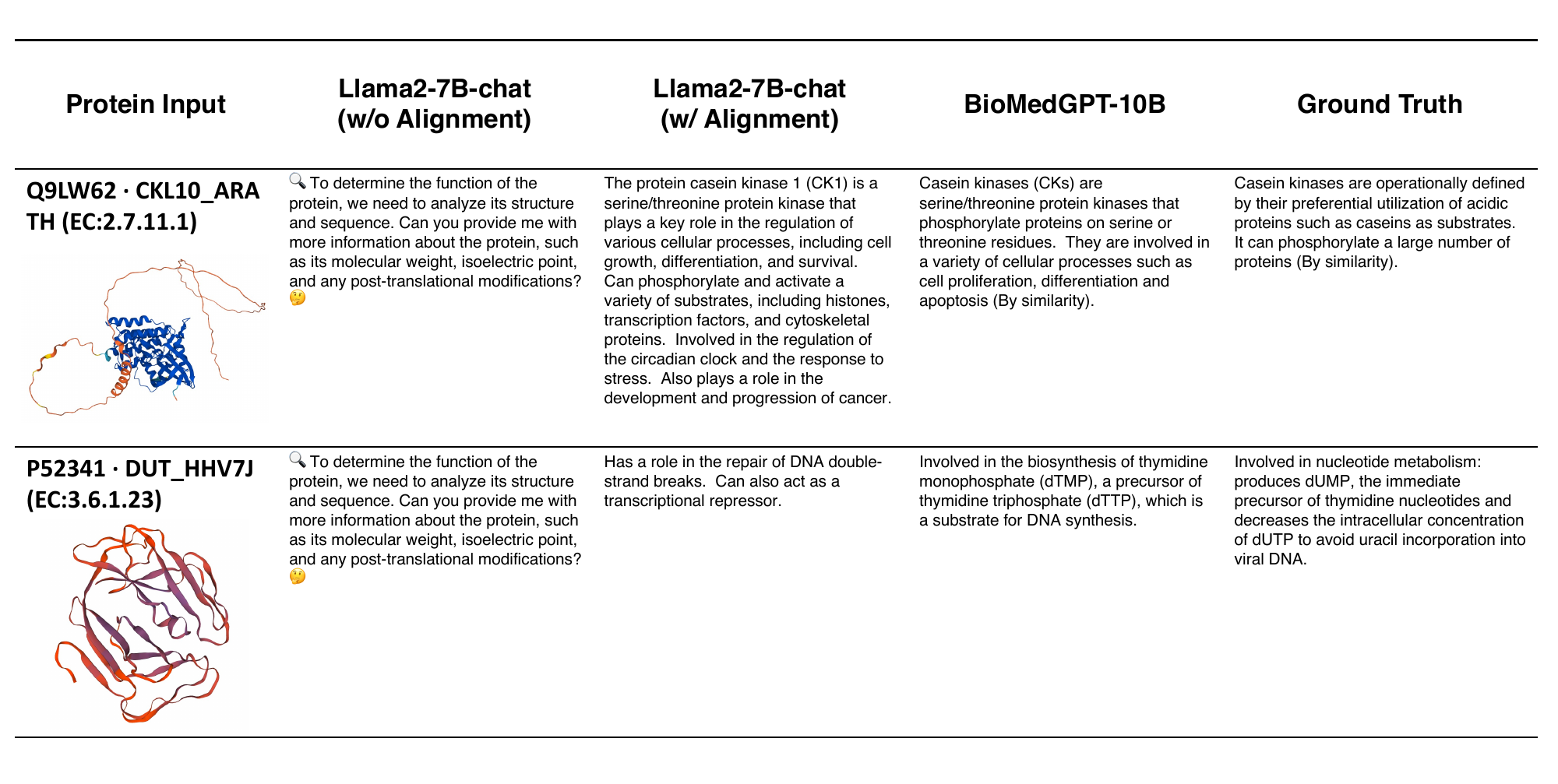}}
  \captionsetup{font={small}}
  \caption{Cases of Protein QA.}
  \label{protein}
\end{figure}

Further, we corroborate our observations with qualitative analysis in Figure \ref{protein}, where we query each model to describe the functions of Q9LW62 · CKL10\_ARATH (EC:2.7.11.1) \citep{Menges2002CellCG} and P52341 · DUT\_HHV7J (EC:3.6.1.23) \citep{Nicholas1996DeterminationAA}. We observe that the original LLama2-7B-chat model yields the same response, asking for more information about the protein. With feature space alignment, both Llama2-7B-Chat and BioMedGPT-10B generates reasonable function annotations. Benefiting from the vertical domain knowledge attained during incremental fine-tuning, BioMedGPT-10B generates more precise descriptions, recognizing serine/threonine phosphorylation's role of casein kinase in cellular processes, as well as thymine nucleotide metabolism's role in nucleotide metabolism.

Establishing a connection between proteins and natural language can provide researchers with richer, more informative, and easily interpretable hints regarding the study of unknown proteins. Furthermore, it can facilitate faster and more accurate protein function annotation.

\section{Limitations}
\textbf{Professional Evaluation}: In this technical report, we predominantly focus on language-centric measures including BLEU and ROUGE scores to evaluate the quality of generated answers, which may be insufficient to gauge the effectiveness of LLMs in the complex biomedical landscape. The intricacies of biomedical data, ranging from molecular interactions to protein functions, demand researchers to develop more nuanced evaluation metrics to capture the accuracy and relevance of biomedical insights provided by LLMs. Furthermore, introducing expert evaluations from domain specialists offers an invaluable perspective in assessing the aptness of LLMs for real-world biomedical applications.



\textbf{Interpretability}: Since LLMs are complex black box models, it is essential to understand the chain of thought behind their generations. Particularly, in assisting biomedical research, the deficiency in interpretability can significantly hamper the ability to provide scientific insights with soundness. 

\textbf{Safety}: While large-scale language models serve as a novel technique, their generated outputs are determined by a certain probability distribution, resulting in unforeseen dangers of generating bias, discrimination, or harmful content. Though we have endeavored to reduce the potential risk of BioMedGPT by fine-tuning on meticulously curated English biomedical corpus, it is hard to fully eliminate this problem. It is essential to ensure the responsible and ethical use of BioMedGPT. While BioMedGPT is endowed with expertise in biomedicine and chemistry, we emphasize that it should \textbf{NOT} be employed for research scenarios that endanger human life, and any further real-world applications should undergo cautious and professional supervision and comprehensive experiments. Users need to be cautious and take extra care when using these models. 


\section{Conclusions and Future Work}
The advent of foundation models has revolutionized the landscape of AI applications across diverse domains, with BioMedGPT standing as a pioneering multi-modal foundation model in the biomedical domain. This work shed light on the potential of leveraging large generative language models to bridge the gap between the language of life and human natural language, empowering a deeper understanding of fundamental life codes in biomedicine. We introduce BioMedGPT-10B, a model instance of BioMedGPT, which leverages the incremental fine-tuning of large language models to comprehend biomedical documents and aligns the feature spaces of molecules, proteins, and natural language. BioMedGPT-10B allows users to easily communicate with various biomedical modalities using free texts. This capability could greatly accelerate the discovery of novel molecules, and therapeutic targets, and empower a wide range of downstream applications in chemistry and biomedicine. We open-source BioMedGPT-10B as well as the fine-tuned language model BioMedGPT-LM-7B to facilitate future research. We are endeavoring to extend the strong capability of BioMedGPT in biomedicine by designing more reliable evaluation metrics, promoting the interpretability of large language models, and enforcing safety measures. Hopefully, BioMedGPT would spark the next generation of biomedical research with human and machine intelligence.




\newpage
\small

\bibliographystyle{unsrtnat}
\bibliography{references}

\begin{thebibliography}{35}
\providecommand{\natexlab}[1]{#1}
\providecommand{\url}[1]{\texttt{#1}}
\expandafter\ifx\csname urlstyle\endcsname\relax
  \providecommand{\doi}[1]{doi: #1}\else
  \providecommand{\doi}{doi: \begingroup \urlstyle{rm}\Url}\fi

\bibitem[OpenAI(2022)]{chatgpt}
OpenAI.
\newblock Introducing chatgpt.
\newblock 2022.
\newblock \doi{https://openai.com/blog/chatgpt}.

\bibitem[Google(2023)]{bard}
Google.
\newblock Bard.
\newblock 2023.
\newblock \doi{https://bard.google.com/}.

\bibitem[Hoffmann et~al.(2022)Hoffmann, Borgeaud, Mensch, and
  et~al.]{Hoffmann2022TrainingCL}
Jordan Hoffmann, Sebastian Borgeaud, Arthur Mensch, and et~al.
\newblock Training compute-optimal large language models.
\newblock \emph{ArXiv}, abs/2203.15556, 2022.
\newblock URL \url{https://api.semanticscholar.org/CorpusID:247778764}.

\bibitem[Touvron et~al.(2023{\natexlab{a}})Touvron, Martin, Stone, Albert,
  Almahairi, Babaei, Bashlykov, Batra, Bhargava, Bhosale, Bikel, Blecher,
  Ferrer, Chen, Cucurull, Esiobu, Fernandes, Fu, Fu, Fuller, Gao, Goswami,
  Goyal, Hartshorn, Hosseini, Hou, Inan, Kardas, Kerkez, Khabsa, Kloumann,
  Korenev, Koura, Lachaux, Lavril, Lee, Liskovich, Lu, Mao, Martinet, Mihaylov,
  Mishra, Molybog, Nie, Poulton, Reizenstein, Rungta, Saladi, Schelten, Silva,
  Smith, Subramanian, Tan, Tang, Taylor, Williams, Kuan, Xu, Yan, Zarov, Zhang,
  Fan, Kambadur, Narang, Rodriguez, Stojnic, Edunov, and
  Scialom]{Touvron2023Llama2O}
Hugo Touvron, Louis Martin, Kevin~R. Stone, Peter Albert, Amjad Almahairi,
  Yasmine Babaei, Nikolay Bashlykov, Soumya Batra, Prajjwal Bhargava, Shruti
  Bhosale, Daniel~M. Bikel, Lukas Blecher, Cristian~Canton Ferrer, Moya Chen,
  Guillem Cucurull, David Esiobu, Jude Fernandes, Jeremy Fu, Wenyin Fu, Brian
  Fuller, Cynthia Gao, Vedanuj Goswami, Naman Goyal, Anthony~S. Hartshorn,
  Saghar Hosseini, Rui Hou, Hakan Inan, Marcin Kardas, Viktor Kerkez, Madian
  Khabsa, Isabel~M. Kloumann, A.~V. Korenev, Punit~Singh Koura, Marie-Anne
  Lachaux, Thibaut Lavril, Jenya Lee, Diana Liskovich, Yinghai Lu, Yuning Mao,
  Xavier Martinet, Todor Mihaylov, Pushkar Mishra, Igor Molybog, Yixin Nie,
  Andrew Poulton, Jeremy Reizenstein, Rashi Rungta, Kalyan Saladi, Alan
  Schelten, Ruan Silva, Eric~Michael Smith, R.~Subramanian, Xia Tan, Binh Tang,
  Ross Taylor, Adina Williams, Jian~Xiang Kuan, Puxin Xu, Zhengxu Yan, Iliyan
  Zarov, Yuchen Zhang, Angela Fan, Melanie Kambadur, Sharan Narang, Aurelien
  Rodriguez, Robert Stojnic, Sergey Edunov, and Thomas Scialom.
\newblock Llama 2: Open foundation and fine-tuned chat models.
\newblock \emph{ArXiv}, abs/2307.09288, 2023{\natexlab{a}}.
\newblock URL \url{https://api.semanticscholar.org/CorpusID:259950998}.

\bibitem[Wu et~al.(2023{\natexlab{a}})Wu, Irsoy, Lu, Dabravolski, Dredze,
  Gehrmann, Kambadur, Rosenberg, and Mann]{bloomberggPT}
Shijie Wu, Ozan Irsoy, Steven Lu, Vadim Dabravolski, Mark Dredze, Sebastian
  Gehrmann, Prabhanjan Kambadur, David Rosenberg, and Gideon Mann.
\newblock Bloomberggpt: A large language model for finance.
\newblock \emph{ArXiv}, abs/2303.17564, 2023{\natexlab{a}}.
\newblock URL \url{https://api.semanticscholar.org/CorpusID:257833842}.

\bibitem[Cui et~al.(2023)Cui, Li, Yan, Chen, and Yuan]{cui2023chatlaw}
Jiaxi Cui, Zongjian Li, Yang Yan, Bohua Chen, and Li~Yuan.
\newblock Chatlaw: Open-source legal large language model with integrated
  external knowledge bases.
\newblock \emph{arXiv:2306.16092}, 2023.

\bibitem[Gao et~al.(2020)Gao, Biderman, Black, Golding, Hoppe, Foster, Phang,
  He, Thite, Nabeshima, Presser, and Leahy]{gao2020pile}
Leo Gao, Stella Biderman, Sid Black, Laurence Golding, Travis Hoppe, Charles
  Foster, Jason Phang, Horace He, Anish Thite, Noa Nabeshima, Shawn Presser,
  and Connor Leahy.
\newblock The pile: An 800gb dataset of diverse text for language modeling.
\newblock \emph{arXiv:2101.00027}, 2020.

\bibitem[Wu et~al.(2023{\natexlab{b}})Wu, Zhang, Zhang, Wang, and Xie]{pmc}
Chaoyi Wu, Xiaoman Zhang, Ya~Zhang, Yanfeng Wang, and Weidi Xie.
\newblock {PMC-LLAMA}: Further finetuning llama on medical papers.
\newblock \emph{arXiv:2304.14454}, 2023{\natexlab{b}}.

\bibitem[Guo et~al.(2023)Guo, Thost, Song, Balachandran, Das, Chen, and
  Matusik]{pmlr-v202-guo23h}
Minghao Guo, Veronika Thost, Samuel~W Song, Adithya Balachandran, Payel Das,
  Jie Chen, and Wojciech Matusik.
\newblock Hierarchical grammar-induced geometry for data-efficient molecular
  property prediction.
\newblock In \emph{Proceedings of the 40th International Conference on Machine
  Learning}, Proceedings of Machine Learning Research, pages 12055--12076.
  PMLR, 23--29 Jul 2023.
\newblock URL \url{https://proceedings.mlr.press/v202/guo23h.html}.

\bibitem[Xu et~al.(2023)Xu, Woicik, Poon, Altman, and Wang]{Xu2023}
Hanwen Xu, Addie Woicik, Hoifung Poon, Russ~B. Altman, and Shen Wang.
\newblock Multilingual translation for zero-shot biomedical classification
  using biotranslator.
\newblock \emph{Nature Communications}, 2023.

\bibitem[Gu et~al.(2020)Gu, Tinn, Cheng, Lucas, Usuyama, Liu, Naumann, Gao, and
  Poon]{pubmedbert}
Yu~Gu, Robert Tinn, Hao Cheng, Michael Lucas, Naoto Usuyama, Xiaodong Liu,
  Tristan Naumann, Jianfeng Gao, and Hoifung Poon.
\newblock Domain-specific language model pretraining for biomedical natural
  language processing.
\newblock 2020.

\bibitem[Consortium(2022)]{UniProt}
The~UniProt Consortium.
\newblock {UniProt: the Universal Protein Knowledgebase in 2023}.
\newblock \emph{Nucleic Acids Research}, 51\penalty0 (D1):\penalty0 D523--D531,
  11 2022.
\newblock ISSN 0305-1048.
\newblock \doi{10.1093/nar/gkac1052}.
\newblock URL \url{https://doi.org/10.1093/nar/gkac1052}.

\bibitem[Lin et~al.(2022)Lin, Akin, Rao, Hie, Zhu, Lu, Smetanin, dos
  Santos~Costa, Fazel-Zarandi, Sercu, Candido, et~al.]{ESM2}
Zeming Lin, Halil Akin, Roshan Rao, Brian Hie, Zhongkai Zhu, Wenting Lu, Nikita
  Smetanin, Allan dos Santos~Costa, Maryam Fazel-Zarandi, Tom Sercu, Sal
  Candido, et~al.
\newblock Language models of protein sequences at the scale of evolution enable
  accurate structure prediction.
\newblock \emph{bioRxiv}, 2022.

\bibitem[OpenAI(2023)]{openai2023gpt4}
OpenAI.
\newblock Gpt-4 technical report.
\newblock \emph{arXiv:2303.08774}, 2023.

\bibitem[Luo et~al.(2023)Luo, Yang, Hong, Liu, and Nie]{molfm}
Yizhen Luo, Kai Yang, Massimo Hong, Xing~Yi Liu, and Zaiqing Nie.
\newblock {MolFM}: A multimodal molecular foundation model.
\newblock \emph{arXiv:2307.09484}, 2023.

\bibitem[Liu et~al.(2022)Liu, Wang, Liu, Lasenby, Guo, and Tang]{GraphMVP}
Shengchao Liu, Hanchen Wang, Weiyang Liu, Joan Lasenby, Hongyu Guo, and Jian
  Tang.
\newblock Pre-training molecular graph representation with 3d geometry.
\newblock In \emph{International Conference on Learning Representations}, 2022.
\newblock URL \url{https://openreview.net/forum?id=xQUe1pOKPam}.

\bibitem[Lo et~al.(2020)Lo, Wang, Neumann, Kinney, and
  Weld]{lo-wang-2020-s2orc}
Kyle Lo, Lucy~Lu Wang, Mark Neumann, Rodney Kinney, and Daniel Weld.
\newblock {S}2{ORC}: The semantic scholar open research corpus.
\newblock In \emph{Proceedings of the 58th Annual Meeting of the Association
  for Computational Linguistics}, pages 4969--4983, Online, July 2020.
  Association for Computational Linguistics.
\newblock \doi{10.18653/v1/2020.acl-main.447}.
\newblock URL \url{https://www.aclweb.org/anthology/2020.acl-main.447}.

\bibitem[Chen et~al.(2016)Chen, Xu, Zhang, and Guestrin]{chen2016training}
Tianqi Chen, Bing Xu, Chiyuan Zhang, and Carlos Guestrin.
\newblock Training deep nets with sublinear memory cost.
\newblock \emph{arXiv:1604.06174}, 2016.

\bibitem[Dao et~al.(2022)Dao, Fu, Ermon, Rudra, and Ré]{dao2022flashattention}
Tri Dao, Daniel~Y. Fu, Stefano Ermon, Atri Rudra, and Christopher Ré.
\newblock Flashattention: Fast and memory-efficient exact attention with
  io-awareness, 2022.

\bibitem[Xu et~al.(2018)Xu, Hu, Leskovec, and Jegelka]{Xu2018HowPA}
Keyulu Xu, Weihua Hu, Jure Leskovec, and Stefanie Jegelka.
\newblock How powerful are graph neural networks?
\newblock \emph{ArXiv}, abs/1810.00826, 2018.
\newblock URL \url{https://api.semanticscholar.org/CorpusID:52895589}.

\bibitem[Vaswani et~al.(2023)Vaswani, Shazeer, Parmar, Uszkoreit, Jones, Gomez,
  Kaiser, and Polosukhin]{vaswani2023attention}
Ashish Vaswani, Noam Shazeer, Niki Parmar, Jakob Uszkoreit, Llion Jones,
  Aidan~N. Gomez, Lukasz Kaiser, and Illia Polosukhin.
\newblock Attention is all you need.
\newblock \emph{arXiv:1706.03762}, 2023.

\bibitem[Kim et~al.(2022)Kim, Chen, Cheng, Gindulyte, He, He, Li, Shoemaker,
  Thiessen, Yu, Zaslavsky, Zhang, and Bolton]{PubChem}
Sunghwan Kim, Jie Chen, Tiejun Cheng, Asta Gindulyte, Jia He, Siqian He,
  Qingliang Li, Benjamin~A Shoemaker, Paul~A Thiessen, Bo~Yu, Leonid Zaslavsky,
  Jian Zhang, and Evan~E Bolton.
\newblock {PubChem 2023 update}.
\newblock \emph{Nucleic Acids Research}, 51\penalty0 (D1):\penalty0
  D1373--D1380, 10 2022.
\newblock ISSN 0305-1048.
\newblock \doi{10.1093/nar/gkac956}.
\newblock URL \url{https://doi.org/10.1093/nar/gkac956}.

\bibitem[Landrum et~al.(2021)Landrum, Tosco, Kelley, sriniker, Ric, gedeck,
  Vianello, NadineSchneider, Dalke, Kawashima, Dan, Cole, Swain, Turk,
  Cosgrove, AlexanderSavelyev, Vaucher, Jones, W{\'o}jcikowski, Probst,
  guillaume godin, Scalfani, Pahl, Berenger, JLVarjo, strets, DoliathGavid,
  Sforna, and Jensen]{rdkit}
Greg Landrum, Paolo Tosco, Brian Kelley, sriniker, Ric, gedeck, Riccardo
  Vianello, NadineSchneider, Andrew Dalke, Eisuke Kawashima, N~Dan, Brian Cole,
  Matt Swain, Samo Turk, David~O. Cosgrove, AlexanderSavelyev, Alain~C.
  Vaucher, Gareth Jones, Maciej W{\'o}jcikowski, Daniela Probst, guillaume
  godin, Vincent~F. Scalfani, Axel Pahl, Francois Berenger, JLVarjo, strets,
  DoliathGavid, Gianluca Sforna, and Jan~Holst Jensen.
\newblock rdkit/rdkit: 2021\_03\_4 (q1 2021) release.
\newblock 2021.
\newblock URL \url{https://api.semanticscholar.org/CorpusID:244961525}.

\bibitem[Ye et~al.(2023)Ye, Xu, Xu, Ye, Yan, Zhou, Wang, Hu, Shi, Shi, Li, Xu,
  Chen, Tian, Qi, Zhang, and Huang]{ye2023mplugowl}
Qinghao Ye, Haiyang Xu, Guohai Xu, Jiabo Ye, Ming Yan, Yiyang Zhou, Junyang
  Wang, Anwen Hu, Pengcheng Shi, Yaya Shi, Chenliang Li, Yuanhong Xu, Hehong
  Chen, Junfeng Tian, Qian Qi, Ji~Zhang, and Fei Huang.
\newblock mplug-owl: Modularization empowers large language models with
  multimodality, 2023.

\bibitem[Jin et~al.(2021)Jin, Pan, Oufattole, Weng, Fang, and Szolovits]{usmle}
Di~Jin, Eileen Pan, Nassim Oufattole, Wei-Hung Weng, Hanyi Fang, and Peter
  Szolovits.
\newblock What disease does this patient have? a large-scale open domain
  question answering dataset from medical exams.
\newblock \emph{Applied Sciences}, 11\penalty0 (14):\penalty0 6421, 2021.

\bibitem[Pal et~al.(2022)Pal, Umapathi, and Sankarasubbu]{medmcqa}
Ankit Pal, Logesh~Kumar Umapathi, and Malaikannan Sankarasubbu.
\newblock {MedMCQA}: A large-scale multi-subject multi-choice dataset for
  medical domain question answering.
\newblock In \emph{Conference on Health, Inference, and Learning}, pages
  248--260. PMLR, 2022.

\bibitem[Jin et~al.(2019)Jin, Dhingra, Liu, Cohen, and Lu]{pubmedqa}
Qiao Jin, Bhuwan Dhingra, Zhengping Liu, William~W Cohen, and Xinghua Lu.
\newblock {PubMedQA}: A dataset for biomedical research question answering.
\newblock \emph{arXiv:1909.06146}, 2019.

\bibitem[Ouyang et~al.(2022)Ouyang, Wu, Jiang, Almeida, Wainwright, Mishkin,
  Zhang, Agarwal, Slama, Ray, Schulman, Hilton, Kelton, Miller, Simens, Askell,
  Welinder, Christiano, Leike, and Lowe]{ouyang2022training}
Long Ouyang, Jeff Wu, Xu~Jiang, Diogo Almeida, Carroll~L. Wainwright, Pamela
  Mishkin, Chong Zhang, Sandhini Agarwal, Katarina Slama, Alex Ray, John
  Schulman, Jacob Hilton, Fraser Kelton, Luke~E. Miller, Maddie Simens, Amanda
  Askell, Peter Welinder, Paul~Francis Christiano, Jan Leike, and Ryan~J. Lowe.
\newblock Training language models to follow instructions with human feedback.
\newblock \emph{ArXiv}, abs/2203.02155, 2022.
\newblock URL \url{https://api.semanticscholar.org/CorpusID:246426909}.

\bibitem[Touvron et~al.(2023{\natexlab{b}})Touvron, Lavril, Izacard, Martinet,
  Lachaux, Lacroix, Rozi{\`e}re, Goyal, Hambro, Azhar, Rodriguez, Joulin,
  Grave, and Lample]{touvron2023llama}
Hugo Touvron, Thibaut Lavril, Gautier Izacard, Xavier Martinet, Marie-Anne
  Lachaux, Timoth{\'e}e Lacroix, Baptiste Rozi{\`e}re, Naman Goyal, Eric
  Hambro, Faisal Azhar, Aurelien Rodriguez, Armand Joulin, Edouard Grave, and
  Guillaume Lample.
\newblock Llama: Open and efficient foundation language models.
\newblock \emph{ArXiv}, abs/2302.13971, 2023{\natexlab{b}}.
\newblock URL \url{https://api.semanticscholar.org/CorpusID:257219404}.

\bibitem[Diao et~al.(2023)Diao, Pan, Dong, Shum, Zhang, Xiong, and
  Zhang]{lmflow}
Shizhe Diao, Rui Pan, Hanze Dong, KaShun Shum, Jipeng Zhang, Wei Xiong, and
  Tong Zhang.
\newblock Lmflow: An extensible toolkit for finetuning and inference of large
  foundation models.
\newblock \emph{arXiv:2306.12420}, 2023.

\bibitem[Edwards et~al.(2021)Edwards, Zhai, and Ji]{edwards-etal-2021-text2mol}
Carl Edwards, ChengXiang Zhai, and Heng Ji.
\newblock {T}ext2{M}ol: Cross-modal molecule retrieval with natural language
  queries.
\newblock In \emph{Proceedings of the 2021 Conference on Empirical Methods in
  Natural Language Processing}, pages 595--607, Online and Punta Cana,
  Dominican Republic, November 2021. Association for Computational Linguistics.
\newblock \doi{10.18653/v1/2021.emnlp-main.47}.
\newblock URL \url{https://aclanthology.org/2021.emnlp-main.47}.

\bibitem[Edwards et~al.(2022)Edwards, Lai, Ros, Honke, Cho, and Ji]{MolT5}
Carl Edwards, Tuan Lai, Kevin Ros, Garrett Honke, Kyunghyun Cho, and Heng Ji.
\newblock Translation between molecules and natural language.
\newblock In \emph{Proceedings of the 2022 Conference on Empirical Methods in
  Natural Language Processing}, pages 375--413, Abu Dhabi, United Arab
  Emirates, December 2022. Association for Computational Linguistics.
\newblock \doi{10.18653/v1/2022.emnlp-main.26}.
\newblock URL \url{https://aclanthology.org/2022.emnlp-main.26}.

\bibitem[Li et~al.(2023)Li, Liu, Fan, Wei, Liu, Tang, and Li]{li2023empowering}
Jiatong Li, Yunqing Liu, Wenqi Fan, Xiao-Yong Wei, Hui Liu, Jiliang Tang, and
  Qing Li.
\newblock Empowering molecule discovery for molecule-caption translation with
  large language models: A chatgpt perspective.
\newblock \emph{arXiv:2306.06615}, 2023.

\bibitem[Menges et~al.(2002)Menges, Hennig, Gruissem, and
  Murray]{Menges2002CellCG}
Margit Menges, Lars Hennig, Wilhelm Gruissem, and James A.~H. Murray.
\newblock Cell cycle-regulated gene expression inarabidopsis *.
\newblock \emph{The Journal of Biological Chemistry}, 277:\penalty0 41987 --
  42002, 2002.
\newblock URL \url{https://api.semanticscholar.org/CorpusID:14605854}.

\bibitem[Nicholas(1996)]{Nicholas1996DeterminationAA}
John Nicholas.
\newblock Determination and analysis of the complete nucleotide sequence of
  human herpesvirus.
\newblock \emph{Journal of Virology}, 70:\penalty0 5975 -- 5989, 1996.
\newblock URL \url{https://api.semanticscholar.org/CorpusID:12112753}.

\end{thebibliography}

\end{document}